\documentclass[
superscriptaddress,
showpacs,preprintnumbers,
amsmath,amssymb,
prl,
twocolumn,
notitlepage
]{revtex4-1}
\usepackage{microtype}
\usepackage{graphicx}
\usepackage{import}
\usepackage{hyperref}
\usepackage{braket}
\usepackage{siunitx}
\usepackage{nicefrac}
\usepackage{changes}
\usepackage{chemformula}
\newcommand*{\Alox}{\ch{AlO_x}}

\newcommand{\Es}{$E_\mathrm{s} $}
\newcommand{\EL}{$E_\mathrm{L} $}
\newcommand{\Rn}{$R_\mathrm{n} $}
\newcommand{\Vc}{$V_\mathrm{c} $}
\newcommand{\Ic}{$I_\mathrm{c} $}
\begin{document}
	
	\title{Insulating, metallic and superconducting behavior in a single nanowire}
	%-----------------------------------------------------------------------------
	\author{Jan Nicolas Voss}
	\affiliation{Physikalisches Institut, Karlsruher Institut für Technologie, 76131 Karlsruhe, Germany}
	\author{Yannick Sch\"on}
	\affiliation{Physikalisches Institut, Karlsruher Institut für Technologie, 76131 Karlsruhe, Germany}
	\author{Micha Wildermuth}
	\affiliation{Physikalisches Institut, Karlsruher Institut für Technologie, 76131 Karlsruhe, Germany}
	\author{Dominik Dorer}
	\affiliation{Physikalisches Institut, Karlsruher Institut für Technologie, 76131 Karlsruhe, Germany}
	\author{Jared H. Cole}
	\affiliation{Chemical and Quantum Physics, School of Science, RMIT University, Melbourne, Victoria 3000, Australia}
	\author{Hannes Rotzinger}
	\email{hannes.rotzinger@kit.edu}
	\affiliation{Physikalisches Institut, Karlsruher Institut für Technologie, 76131 Karlsruhe, Germany}
	\affiliation{Institute for Quantum Materials and Technologies, Karlsruher Institut für Technologie, 76021 Karlsruhe, Germany}
	\author{Alexey V. Ustinov}
	\affiliation{Physikalisches Institut, Karlsruher Institut für Technologie, 76131 Karlsruhe, Germany}
	\affiliation{National University of Science and Technology MISIS, Moscow 119049, Russia}
	\affiliation{Russian Quantum Center, Skolkovo, Moscow 143025, Russia}
	\date{\today}
		
	\begin{abstract} In systems with reduced dimensions quantum fluctuations have a strong influence on the electronic conduction, even at very low temperature. In superconductors this is especially interesting, since the coherent state of the superconducting electrons is strongly interacting with these fluctuations and therefore is a sensitive tool to study them.
	
	In this paper, we report on comprehensive measurements of superconducting nanowires in the quantum phase slip regime. Using an intrinsic electromigration process, we have developed a method to lower the resistance of lithographically fabricated highly resistive nanowires in situ and in small consecutive steps. At low temperature we observe critical (Coulomb) blockade voltages and superconducting critical currents, depending on the nanowire's normal-state resistance, in good agreement with theoretical models. Between these two regimes, we find a continuous transition displaying a nonlinear metallic-like behavior. 
	
	The reported intrinsic electromigration technique is not limited to low temperatures as we find a similar change in resistance that spans over three orders of magnitude also at room temperature. Aside from superconducting quantum circuits, such a technique to reduce the resistance may also have applications in modern electronic circuits. 
	\end{abstract}

\maketitle

When the dimensions of electronic circuits are reduced to nanometer scales their transport characteristics can change fundamentally. A superconducting wire, which is approximately as narrow as the variation length scale of the superconducting order parameter, responds to an applied electrical field in dramatically different ways: The behavior of a short wire with a low normal-state resistance resembles, at temperatures well below the critical temperature, the well-known behavior of a bulk superconductor. The order parameter is well defined and the resistance vanishes. A long wire with a sufficiently high normal-state resistance, instead, does not even reveal superconductivity at a first glance, since the wire does not conduct electrical current at low applied voltages.

This intriguing phenomenon has its origin in the strong confinement of the superconducting condensate which leads to a highly fluctuating order parameter, and therefore its phase 'slips' \cite{Giordano1988,Bezryadin2000,Nazarov2011,Astafiev2012} (for an overview of the effect see e.g.\@ \cite{ARUTYUNOV20081}). At temperatures close to the transition temperature, the electrical response is governed by thermally activated phase slips \cite{TAPS1,TAPS2}. At very low temperatures, however, the origin of these phase slips has a quantum nature. 
Between the two extreme states, superconducting and insulating, the response shows a nonlinear metallic-like, behavior at small applied voltages (in the following for simplicity denoted as 'metallic').

Until now, it was not possible to access these three different regimes at low magnetic field with a single wire \cite{Kim, baumans_thermal_2016, Koval}, since the intrinsic properties, like coherence length or nanowire resistance, were fixed by the preparation of the wire.

\begin{figure}[h!]	
	\begin{minipage}[l]{0.605\columnwidth}
		\includegraphics[width=\textwidth]{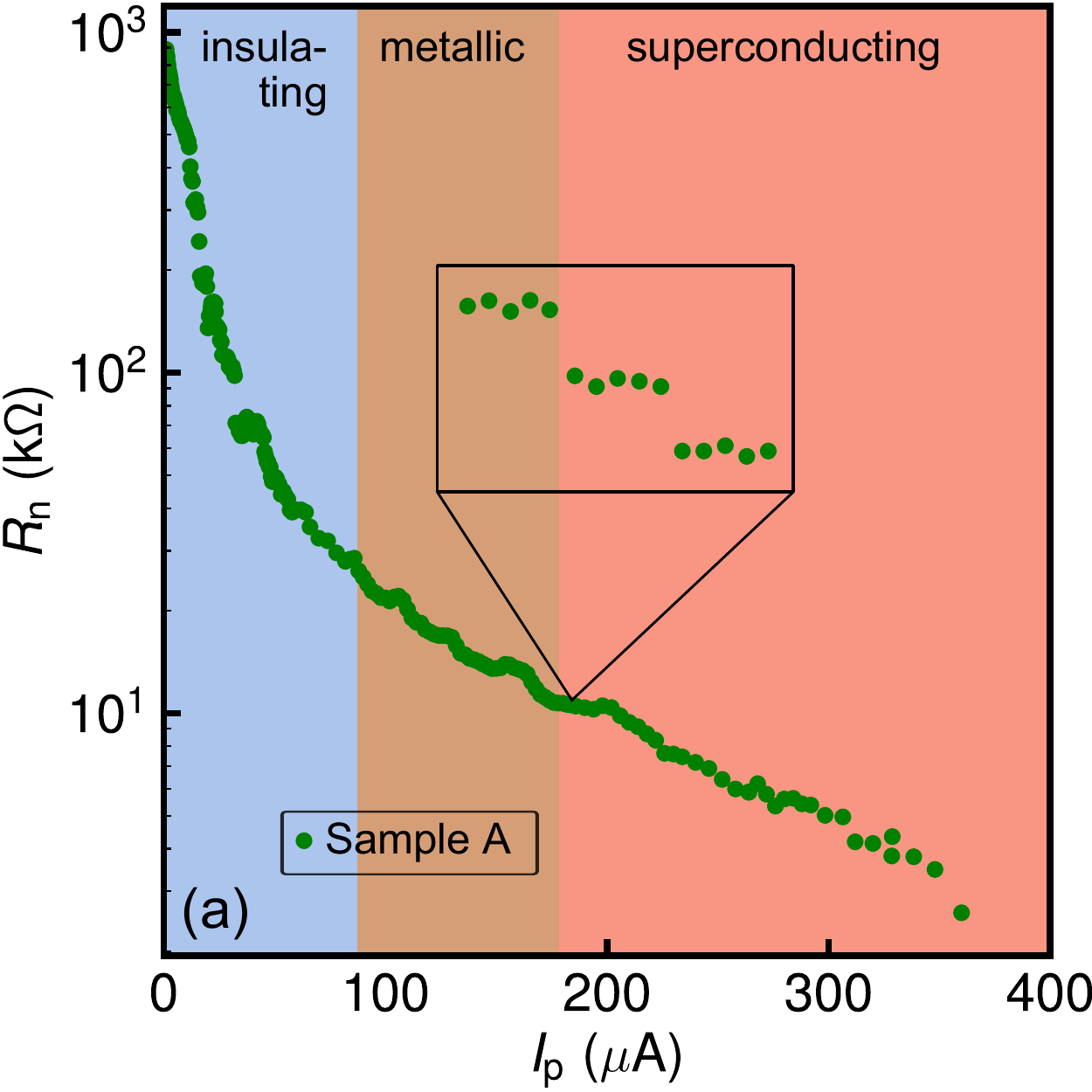}
	\end{minipage}	
	\begin{minipage}[r]{0.38\columnwidth}
		\includegraphics[width=\textwidth]{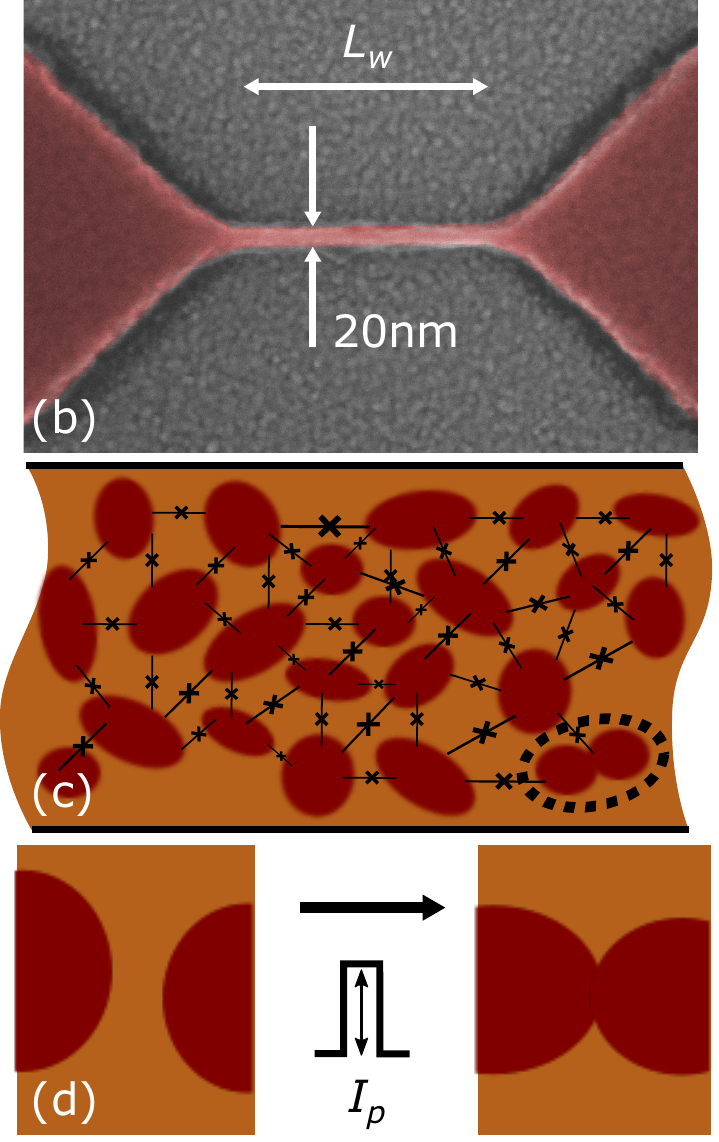}
	\end{minipage}
	\hfill	
	\caption{\textbf{(a)} Normal state resistance of an oxidized (granular) aluminum nanowire as a function of the applied pulse current $I_\mathrm{p}$ measured at 25\,mK. The resistance of the 20\,nm wide and 1000\,nm long nanowire is stepwise (inset) lowered from 900\,k$\Omega$ to 2.5\,k$\Omega$. At low bias values an insulating, metallic and superconducting behavior is observed. \textbf{(b)} Scanning electron micrograph of a lithographically fabricated nanowire (coloured in red) on a sapphire substrate. \textbf{(c)} Illustration of the wire structure. Isolated aluminum grains (red, size $\simeq$ 4\,nm) are embedded in an aluminum-oxide matrix (orange), forming a network of tunnel junctions (black crosses). \textbf{(d)} Proposed microscopic process: The current pulses lead to an intrinsic electromigration process (IEM) merging grains and / or clusters of grains.}
	
	\label{schematic}
\end{figure}

In this paper, we present a technique which allows for the permanent reduction of the normal-state resistance of single wires by three orders of magnitude compared to its initial resistance at low temperatures. We use current pulses of increasing amplitude to alter the internal structure of oxidized (granular) aluminum (\Alox) nanowires by an intrinsic electromigration (IEM) process. By doing so, we observe a transition of the electrical response from the insulating through the metallic to the superconducting state. 

The studied thin films consist of a network of small aluminum grains with a diameter of about 4\,nm covered by a thin aluminum-oxide insulator (see TEM picture in supplementary material of Ref.\@ \cite{Rotzinger2016}). This inter-grain matrix plays a dominant role in the transport properties of the material and, at very low temperatures where the aluminum is in the superconducting state, is usually described as a network of Josephson weak links. The charge transport between the grains and thus also the normal-state resistance are therefore strongly dependent on the thickness of the insulating barriers. In the fabrication process of the films, the thickness of the insulating layer between the grains can be influenced by the amount of oxygen added to the deposited pure aluminum. These films can have a normal sheet resistance up to several k$\Omega$ and exhibit low intrinsic microwave losses in the superconducting state, making them a versatile material for high impedance superconducting quantum circuits (see e.g.\@ \cite{Rotzinger2016, Maleeva2018, Grunhaupt2018, Schon2019}). The kinetic inductance $L_\mathrm{k}$ of a wire made of such film may exceed the geometric inductance by orders of magnitude. It is convenient to describe it by $L_\mathrm{k} = \hbar R_\mathrm{n}/ \pi \Delta = 0.18\, \hbar R_\mathrm{n}/k_\mathrm{B} T_\mathrm{c}$, where $R_\mathrm{n}$, $\Delta$, $T_\mathrm{c}$ are the normal-state resistance, the (BCS) superconducting energy gap and the transition temperature ($1.4- 2\,K$ for granular aluminum)\cite{Rotzinger2016}.

Due to the Josephson-coupling between the grains, the transport properties of \Alox\ films are inherently non-linear. However, if the number of parallel grains in a wire is large, the Josephson non-linearity is washed out and becomes only visible at large electrical currents. \\

\textbf{Theoretical background }Neglecting the microscopic disorder, the transport dynamics of superconducting nanowires in the insulating regime has been proposed to be dual to the transport dynamics of a Josephson junction in the superconducting regime \cite{Mooij_2006}.
Following \cite{Mooij_2015}, a measure of the strength of the phase fluctuations in a wire is given by the phase slip energy 
 \begin{equation}
	E_\mathrm{s} = \alpha \left(\frac{ \mathrm{L_w}}{\xi} \right)^2 \mathrm{k_B T_c} \frac{\mathrm{R_q}}{R_\mathrm{n}} \exp \left(-\beta \frac{\mathrm{R_q} \mathrm{L_w}}{\xi R_\mathrm{n}} \right)
	\label{Es}
\end{equation}

where $\mathrm{L_w}$, $\mathrm{\xi}$, $\mathrm{R_q}= h/4e^2$, are the length of the wire, the effective superconducting coherence length and the superconducting resistance quantum, respectively. The empirical constants $\alpha$ and $\beta$ are of the order of 1 \cite{ARUTYUNOV20081}. The phase slip energy exponentially depends on the normal-state resistance \Rn. Thus, \Es\ can vary by orders of magnitude within the \Rn\ range studied in this paper. 

The inductive energy $E_\mathrm{L} = \Phi_0^2/2L_\mathrm{k}$ also plays an important role in the electrical response of a wire. If \EL\ is much larger than \Es, the superconducting phase difference along the wire is well defined showing a superconducting behavior. Here coherent transport of Cooper pairs leads to a vanishing voltage drop up to a critical current $I_\mathrm{c}$; for a review see e.g.\@ \cite{Likharev1979}. In the insulating regime, where \Es\ is much larger than \EL, no conductance is observed up to an applied critical voltage of $V_\mathrm{c} = 2\pi/2e\, E_\mathrm{s}$. When altering \Rn, both \Es\ and \EL\ are changed, \EL\ however, only linearly \cite{Mooij_2015}

\begin{equation}
	\frac{E_\mathrm{s}}{E_\mathrm{L}} = \alpha \left(\frac{\mathrm{L_w}}{\xi} \right)^2 \frac{0.18}{\mathrm{\pi}} \exp \left(-\beta \frac{\mathrm{R_q} \mathrm{L_w}}{\xi R_\mathrm{n}} \right).
	\label{EsEl}
\end{equation}
 
As a result, already small changes in \Rn\ have a high impact on the ratio between $E_\mathrm{s}$ and $E_\mathrm{L}$ and thus on the transport properties of the wire. Following Mooij et al. \cite{Mooij_2015}, the transition from an insulating to a superconducting behavior should happen at \Es\,$\approx$\,0.2 \EL.

\begin{table}
	\caption{\textbf{Parameters of the samples measured at mK temperatures}. All \Alox\ nanowires have a width and a thickness of 20\,nm. $R_{\mathrm{n}}^{0}$, $R_{\mathrm{n}}^{\mathrm{E}}$, denote the initial and final normal-state resistance (before and after altering the nanowire), $E_\mathrm{s}^0$ and $E_\mathrm{s}^\mathrm{E}$ are the corresponding phase slip energies. $R_\mathrm{n}^\mathrm{m}$, $E_\mathrm{s}^\mathrm{m}$ and $R_\mathrm{n}^\mathrm{s}$, $E_\mathrm{s}^\mathrm{s}$ are the largest metallic and superconducting normal-state resistances / phase slip energies.}
	\label{table}
	\begin{center}
		\begin{tabular}{c|c|c|c|c|c|c|c|c|c|c}
			\# & $\mathrm{L_w}$ & $R_\mathrm{n}^{0}$ & $R_\mathrm{n}^{E}$ & $R_\mathrm{n}^\mathrm{m}$ & $R_\mathrm{n}^\mathrm{s}$ & $\frac{R_\mathrm{n}^{0}}{R_\mathrm{n}^\mathrm{E}}$ & $\frac{E_\mathrm{s}^0}{h}$ & $\frac{E_\mathrm{s}^\mathrm{m}}{h}$ &$\frac{E_\mathrm{s}^\mathrm{s}}{h}$ &$\frac{E_\mathrm{s}^\mathrm{E}}{h}$\\
			& nm & k$\Omega$ & k$\Omega$ & k$\Omega$ & k$\Omega$ &  & GHz & GHz & MHz & Hz \\
			\hline
			A & 1000 & 900 & 2.5 & 37 & 16 & 360 & 200 & 2.5 & 1 & 10e-21 \\
			B & 750 & 500 & 3.7 & 28 & 17 & 135 & 164 & 3.0 & 34 & 10e-7\\
			C & 250 & 12.4 & 1.5 & 12.4 & 4.7 & 8.3 & 0.5 & 0.5 & 2.5e-3 & 10e-16\\	\end{tabular}
	\end{center}
\end{table}
 
\section{Results and discussion}

\textbf{Reducing the normal-state resistance} Our experiments focus on a scheme of applying current pulses and measuring the changes in the normal-state resistance \Rn\ of a wire together with the current-voltage ($I-V$) characteristics in the superconducting state. Figure~\ref{schematic} (a) shows a typical resistance \Rn\ vs.\@ pulse amplitude $I_\mathrm{p}$ measurement at 25\,mK here for sample A (length 1000\,nm).  We applied current pulses  with increasing amplitudes ranging from $I_\mathrm{p} \simeq 1\,\mu \mathrm{A}$ to $I_\mathrm{p} = 380\,\mu \mathrm{A}$ in 240 steps. Once a reduction of \Rn\ at a certain threshold current $I_\mathrm{p}$ is observed, applying pulses with an amplitude below the next threshold does not change \Rn. This behavior is illustrated in the inset of Fig.~\ref{schematic} (a). After the resistance is changed to a certain \Rn\, it remains stable, also after thermal cycling of the cryostat to room temperature. In addition, to ensure that the altered resistance values, as well as the transport characteristics, are stable in time, test measurements were performed over days. No recovery of either the normal-state resistance or the transport response was observed. We have applied the described measurement scheme at various temperatures to about 25 nanowire samples of different length and from different fabrication batches, with similar results. Due to the character of the change in resistance, we name the method as intrinsic electromigration (IEM).

We suggest the following microscopical origin for the alteration of the resistances, illustrated in Fig.\ref{schematic} (d). With a current applied to the nanowire, a local voltage develops which mainly drops over the insulating grain-to-grain interlayers. At a threshold current $I_\mathrm{p}$ the tunnel junction is pinched out, merging two or more grains with the weakest insulating barrier \cite{EM_1,EM_2}. We observe that the current pulses can be either applied at room temperature or at low temperatures with very similar results, leading to a permanent change in resistance. It is advantageous, especially at low temperatures, to apply relatively short current or voltage pulses to avoid unnecessary heating. Our findings indicate that this procedure creates a new network of more strongly connected grains, which effectively leads to an increase in the wire conductance, as seen in Fig. \ref{schematic} (a).

\begin{figure}[h!]
	
	\begin{minipage}[c]{0.99\columnwidth}
		\includegraphics[width=\textwidth]{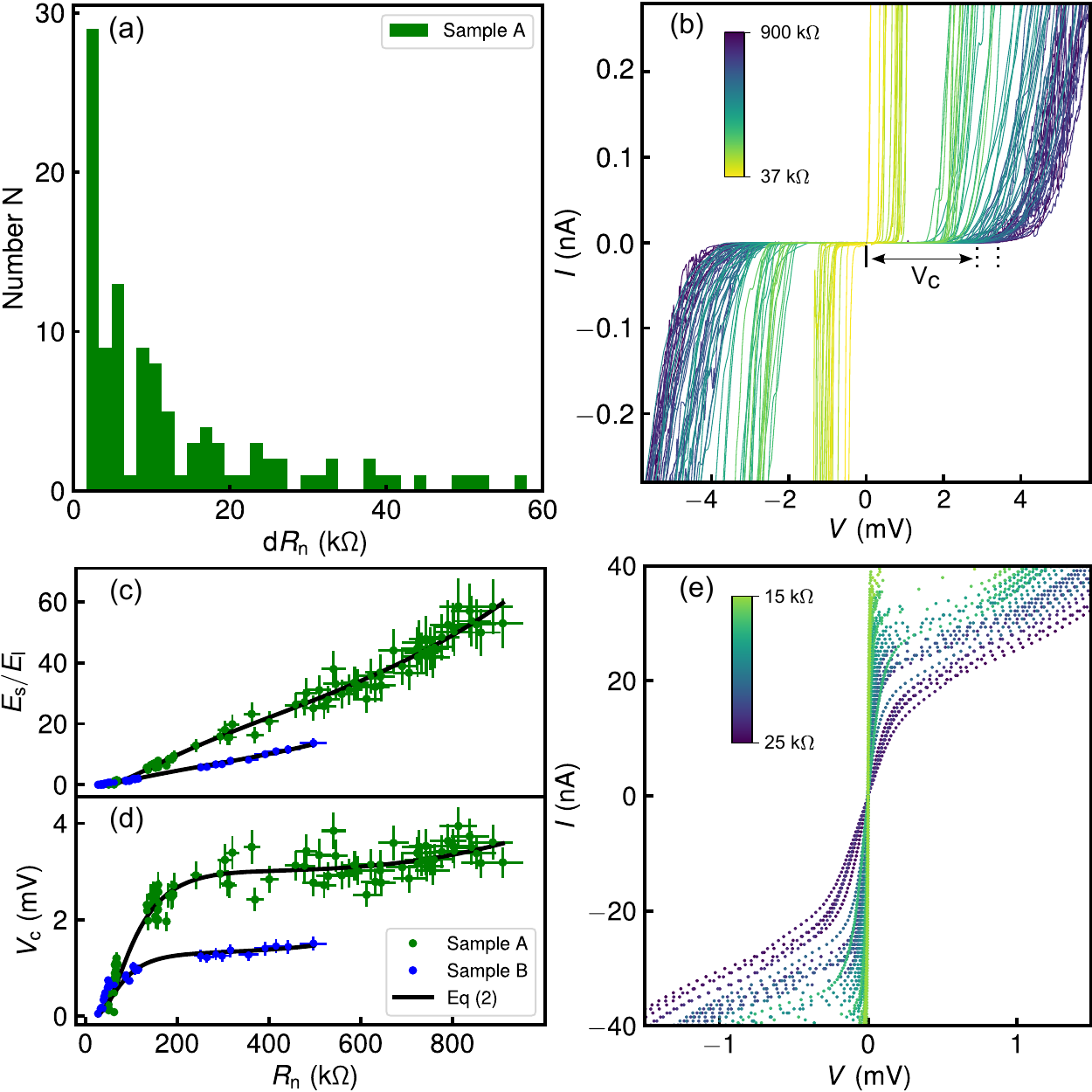}
	\end{minipage}
	\hfill
	\caption{\textbf{(a)} Distribution of resistance steps d\Rn\ for sample A. Steps of d\Rn\ $> 20$\,k$\Omega$ occur only at very small $I_\mathrm{p}$. \label{fig:fofA} \textbf{(b)} $I-V$ characteristics for a 1000 nm long nanowire. Darker curves correspond to higher resistance values, brighter to lower. The Coulomb blockade range is from 3.6\,mV for a normal-state resistance of $\approx$ 800\,k$\Omega$ down to about 0.1\,mV for the lowest resistance value ($\approx$\,40\,k$\Omega$).
		\textbf{(c)} Ratio between phase slip energy $E_\mathrm{s}$ and inductive energy $E_\mathrm{L}$ as a function of the normal-state resistance of wires A and B. For both wires, the ratio converges for smaller resistances towards the same value. Here, the transition from insulating to metallic behavior occurs at $E_\mathrm{s} / E_\mathrm{L}\approx 0.02$. \textbf{(d)} Critical voltages $V_\mathrm{c}$ as a function of the normal-state resistance \Rn. The comparison of measured values and the predictions from Eq.\@ \ref{Es} (b) and Eq.\@ \ref{EsEl} (c) with the fitting parameters $\alpha = 0.07\pm 0.01 $ and $\beta= 0.49 \pm 0.03 $ shows a good agreement for both wires (see black solid lines in (a) and (b)). \label{qps_regime}\textbf{(e)} $I-V$ characteristics of sample A in the metallic regime ($\approx$\,25\,k$\Omega$ -15\,k$\Omega$). In the vicinity of zero bias current, the slope increases when the wire resistance is reduced.}
\end{figure}

We note that both, the magnitude of the \Rn\ changes and the adjusting accuracy, are strongly dependent on the wire length. A qualitative explanation for this behavior may lie in the random distribution of barrier thicknesses. To first order, the number of junctions in the network scales linearly with the length of the nanowire. The probability of having a few very weak internal junctions dominating \Rn\ therefore also increases quickly with the length (see Table \ref{table}). As a consequence, the first changes in resistance are very steep (Fig.~\ref{schematic} (a)). However, with a further reduction of resistance, we observe smaller steps. Figure~\ref{fig:fofA} (a) displays the distribution of resistance steps d\Rn\ for sample A, hosting a few thousand separated aluminium grains. \\

\textbf{Insulating regime }By using the IEM method, the resistance values initially change rapidly (Fig.~\ref{schematic} (a)) and we observe  larger gaps between the measured resistance values (order of 10\,k$\Omega$), visible as larger steps in the distribution tail in Fig.~\ref{fig:fofA} (a). Consequently, also the critical voltage values reflect these gaps in \Rn.

Prior to applying any current pulses to lower \Rn, samples A and B, see Tab. \ref{table}, showed strong Coulomb blockaded behavior with maximal critical voltage $V_\mathrm{c}$\,=\,3.6\,mV (Fig.~\ref{qps_regime} (d)).  For high blockade voltages, we observed a continuous rounding of the $I-V$ characteristics at \Vc\ which lead to a significant error contribution in the determination of \Vc\ values $>$\,2\,mV ($\mathrm{E_\mathrm{s}/h }\approx 150$ GHz in the QPS model). We attribute this rounding to an effect of the relatively low impedance of the environment and therefore elevated temperature due to dissipation in the nanowire \cite{Zorin2012}. However, at lower \Vc\ the $I-V$ characteristics become very steep, but not hysteretic in the current values. 

As an important consequence of the lowering of \Rn\ we expect the granular structure to be altered, forming a network of more and more galvanically connected grains (see Fig.~\ref{schematic} (d)). Due to reduced grain boundary scattering \cite{Ambegaokar92_weak_loc} a longer  mean free path $l_0$ also results in an increased coherence length $\xi_\mathrm{eff} = \sqrt{l_0 \xi_{0}}$. Experimentally, this has also been observed in other granular systems which have been treated by classical electromigration \cite{Bose2006}. The data presented in Fig.~\ref{qps_regime} (c,d) are best fitted assuming a linear $\xi$ span from $\xi = 8$\,nm ($\mathrm{R_n}=900$\,k$\Omega$) to $\xi=12$\,nm (\Rn\,=\,37\,k$\Omega$).
 
Figure~\ref{qps_regime} (d) displays the extracted critical voltages as a function of the altered normal-state resistances for both wires. The overlaid curve is a fit to the measured data using Eq.~\ref{Es}. The two extracted parameters $\alpha = 0.07\pm 0.01 $ and $\beta= 0.49 \pm 0.03 $ are common to the data of sample A and B and in good agreement with the values given in Ref.\@ \cite{Mooij_2015}. Figure~\ref{qps_regime} (c) shows excellent agreement with the QPS theory, Eq.~\ref{EsEl}, for samples A and B. Here we also took into account the change in the inductive energy. The ratio $E_\mathrm{s} / E_\mathrm{L}$ drops almost linearly to values close to unity where we find a transition to the metallic regime.\\

\textbf{Metallic regime }We observe a metallic regime for all samples with \Rn\ values between 40\,k$\Omega$ and 16\,k$\Omega$, characterized by a linear response for small bias values and a non-linear response at larger bias values. For sample A, this is shown in Fig.~\ref{fig:fofA} (e),  see also supplementary material. The exact nature of the transition from insulating to metallic, and ultimately to the superconducting phase, is not currently clear. 

From the perspective of the QPS model, using the parameters evaluated above, in this regime the $I-V$ characteristic is associated  with $\mathrm{E_s}$ in the range between $\mathrm{E_s}/h\approx 3$\,GHz  and  $\mathrm{E_s}/h\approx 30$\,MHz, while $\mathrm{E_L}$ changes between $\mathrm{E_L}/h\approx 150$\,GHz  and $\mathrm{E_L}/h\approx 250$\,GHz. In this intermediate regime, $E\mathrm{_s}$ reduces to values where on average neither the localization of  charges in the wire nor the phase coherence across the wire is dominating.

Several observations in our experiments are consistent with previous work on 2D granular films \cite{KivelsonSpivak2019}, and 1D arrays of Josephson junctions. For a detailed discussion on the I-M-S transition in 2D granular superconductors, see \cite{katsumoto_1995, gantmakher_2010, beloborodov_2007}. More recently studies of 1D chains and 2D arrays of Josephson junctions \cite{fazio_2001} have reported similar behavior with transitions between insulating, metallic (quasiparticle dominated) and Cooper-pair dominated transport, as a function of temperature and magnetic field \cite{vogt_2015, vogt_2016, Cedergren_2015, Cedergren_2017}. The interplay between fundamental energy scales in these systems is similar to the granular aluminum films, and so one would expect similar transport properties.
	
If one considers the model of a network of Josephson junctions (illustrated in Fig.\@  \ref{schematic} (c)), there are two energy scales of interest. The finite charging energy of the grains sets the energy scale associated with localized charges in the system, whereas the Josephson energy sets the energy scale of the delocalization which underpins the superconducting state. One can argue that there should be a transition (even at zero temperature) associated with the crossover from Coulomb dominated to Josephson dominated dynamics, as is typically discussed in the context of Josephson junction arrays \cite{fazio_2001}. As more links are connected by the IEM, the charging energy per grain is reduced eventually allowing conduction pathways to form, resulting in a metallic state.\\

\begin{figure}[h]
	\begin{minipage}[c]{0.49\columnwidth}
		\includegraphics[width=\textwidth]{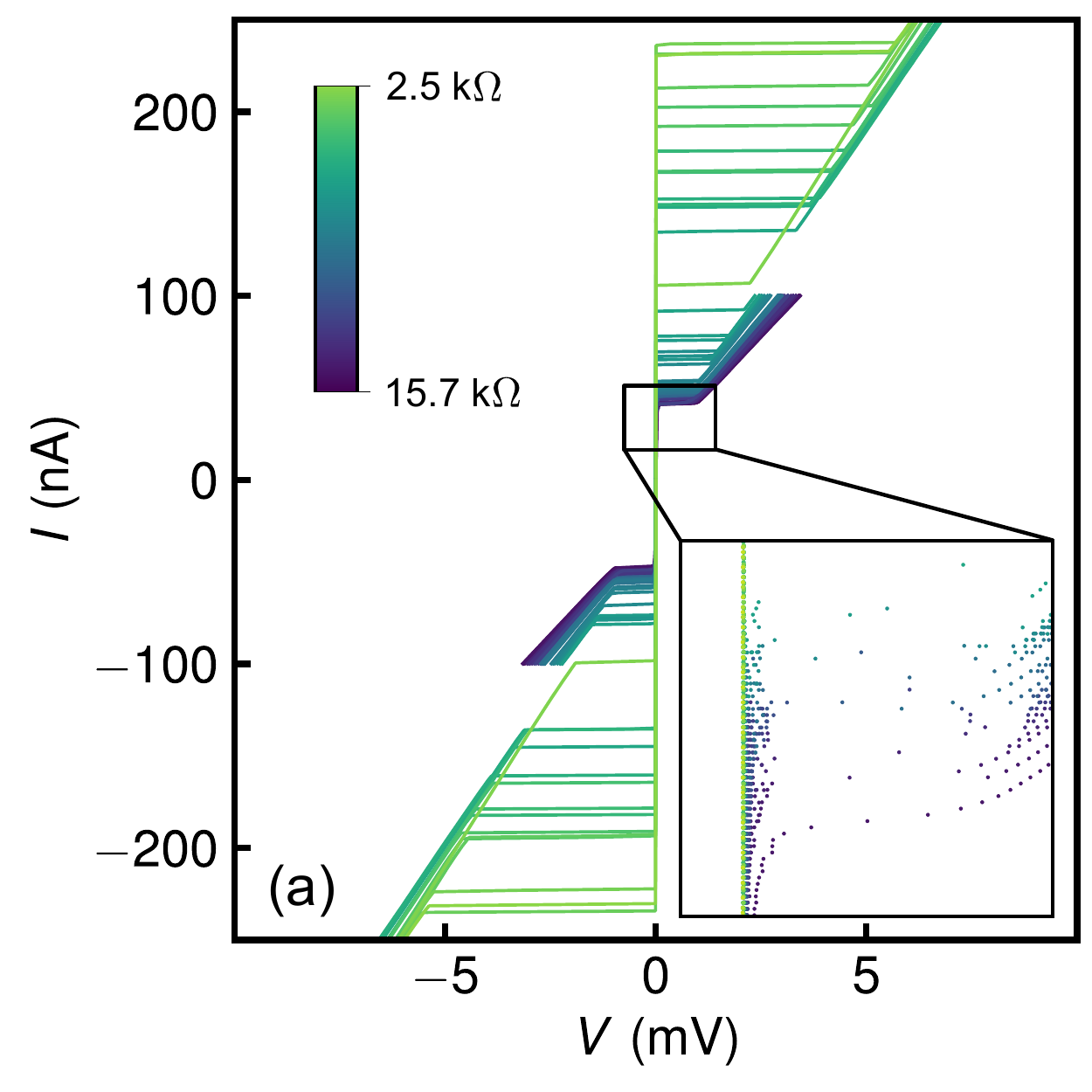}
	\end{minipage}
	\begin{minipage}[c]{0.49\columnwidth}
		\includegraphics[width=\textwidth]{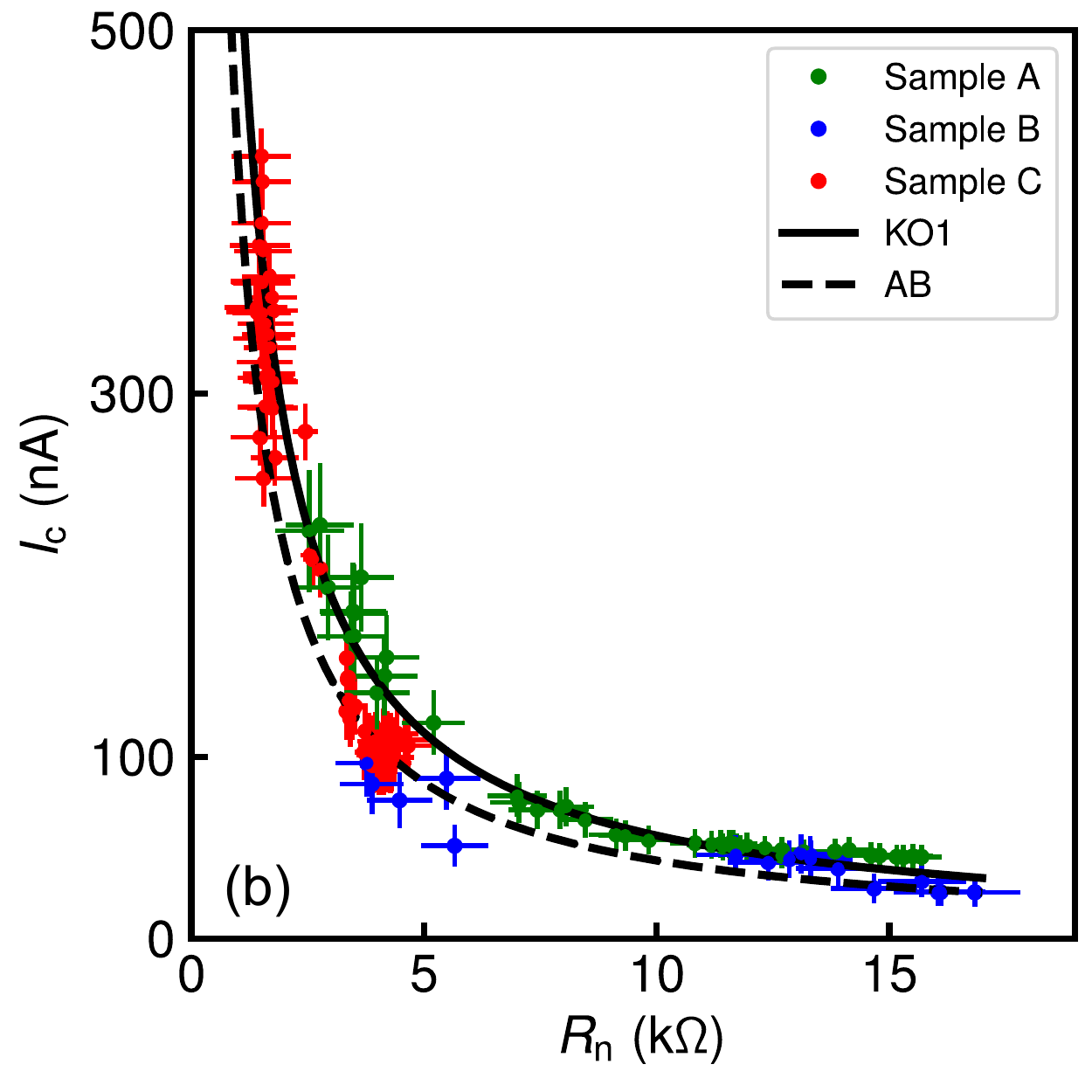} 
	\end{minipage}
	\hfill
	\caption{\textbf{(a)} $I-V$ characteristics of sample A as a function of normal-state resistance. Here \Rn\ is extracted from the resistive slope above $I_\mathrm{c}$ at finite voltages. \textbf{(b)} Critical currents, when the wires are in the superconducting regime, compared with the predicted values from the KO1 (black solid line) and AB (black dashed line) theory. }
	\label{FIG: 1um_supercond}
\end{figure}

\textbf{Superconducting regime }For values of $R\mathrm{_n}/L_\mathrm{W}$ (here $L_\mathrm{W}$ is the wire length) smaller than about 20\,$\Omega/\mathrm{nm}$, the $I-V$ characteristics display a transition into a supercurrent state with no voltage drop up to a critical current \Ic. 
In Fig.~\ref{FIG: 1um_supercond} (a) this is shown for sample A. At the largest \Rn\ and at current values close to \Ic\ the voltage drop across the wire develops rather smoothly (inset), very similar to the phase-diffusion behavior of small capacitance Josephson junctions. Larger \Ic\ values show a voltage discontinuity which develops in magnitude as \Rn\ decreases as it is common for superconducting wires.
 
Under the assumption that the QPS model is still valid and that the above determined empirical parameters $\alpha$ and $\beta$ are unchanged, we can compute \Es\ to be of the order of 1 MHz, while \EL\ can be a few hundred MHz. Due to the smallness of \Es, however, it is more useful to describe the wire as a narrow superconducting filament. Therefore we compare the measured $I_\mathrm{c}$ to a model for short weak links in the dirty limit (mean free path $l_0\ll \mathrm{L_{link}}$) (Kulik and Omel'Yanchuk, KO1) \cite{Kulik1975} and to the expectation that the wire would behave like a Josephson tunnel junction (Ambegaokar and Baratoff, AB) \cite{AB_PhysRevLett.10.486}. For details see supplementary material.

For both models, $I_\mathrm{c}$ as a function of the normal-state resistance is given by:
$\braket{I_\mathrm{c}} = g' ({\pi \bigtriangleup_{\mathrm{BCS}}}/{2e} ) {\braket{R_\mathrm{n}}}^{-1} \label{KO1}$ with $g' = 1.32$ (KO1) or $g' = 1$ (AB). Our measurements do not allow the determination of the current-phase relation of the wire at a given resistance and thus cannot be compared with both models in this respect. Fig.~\ref{FIG: 1um_supercond} (b) shows the calculated values for both models, together with the $I_\mathrm{c}$s from all three samples.

Connecting the picture with Josephson networks, we can interpret the M-S transition as the point at which the average Josephson energy is larger than the Coulomb energy and the usual Bose gas/glass description would now apply. If the superconducting gap is suppressed in the smallest grains \cite{bose_2010}, the fusing of grains will also have the effect of reducing the suppression, increasing the gap and further strengthening the superconducting phase.\\

\begin{figure}
	\begin{minipage}[c]{0.85\columnwidth}
		\includegraphics[width=\textwidth]{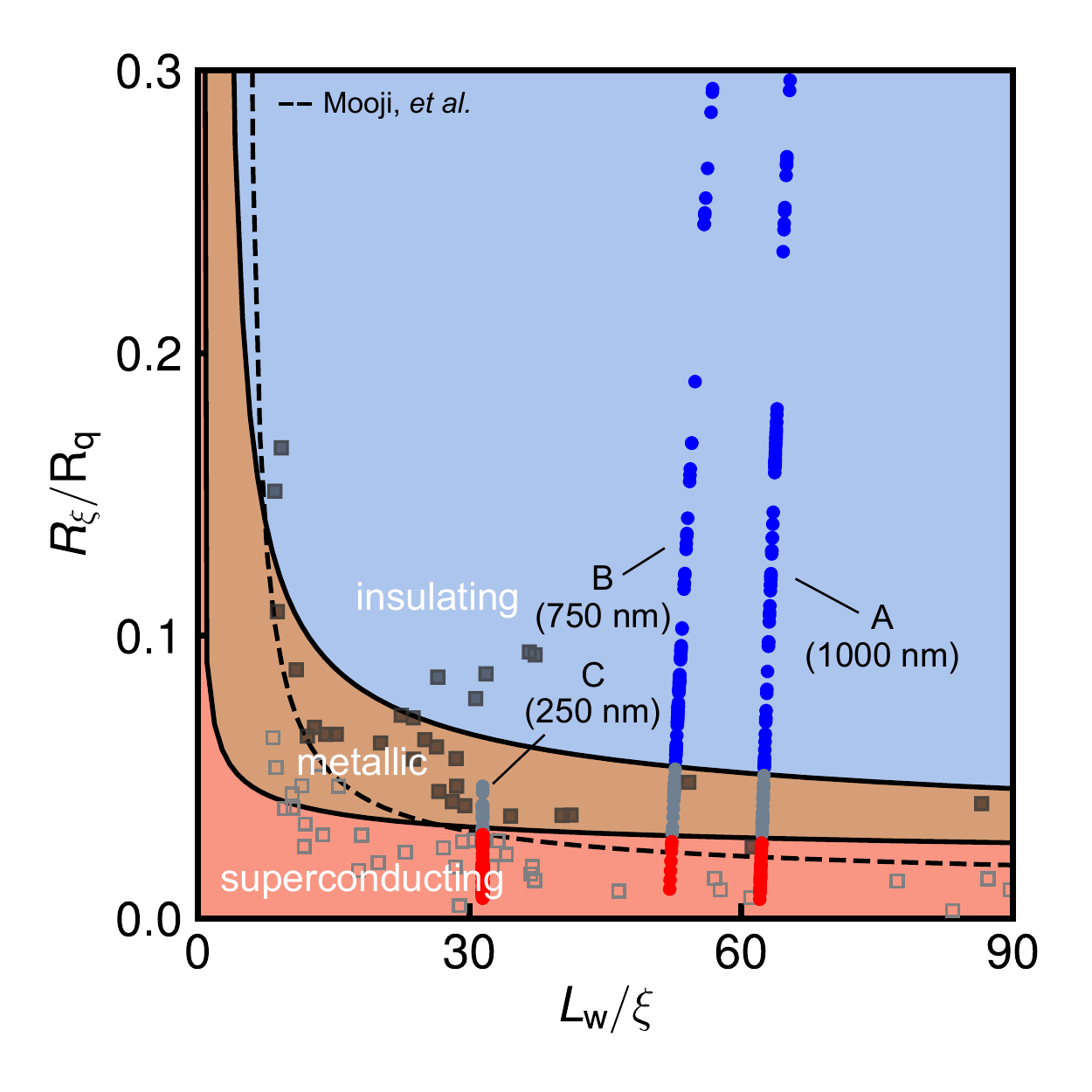}
	\end{minipage}	
	\hfill
	\caption{Tentative phase diagram for nanowires made from oxidized (granular) aluminum. For the black lines, the ratio between the phase slip energy and the inductive energy of the wires is constant for different geometries and specific resistances. The dots represent the altered normal-state resistances and the colour the low temperature state (blue = insulating, gray = metallic, red = superconducting). Assuming a constant wire geometry and only small changes in the coherence length, the ratio between $E_\mathrm{s}$ and $E_\mathrm{L}$ decreases together with $R_\mathrm{\xi}$. From the smallest measurable critical voltages, we find a constant ratio of $E_\mathrm{s} / E_\mathrm{L}\approx 0.02$ (see Fig. \ref{qps_regime} (b)) for the insulating to metallic transition (upper black line). The ratio at which the transition, metallic to superconducting, occurs ($E_\mathrm{s} / E_\mathrm{L}\approx 10^{-4}$, lower black line), is determined from the smallest \Rn\ values in the superconducting regime. The black and gray squares represent the data from Bollinger et al. \cite{Bollinger2008}
\label{phase_diagram}}
\end{figure}

\textbf{Phase diagram }We construct a tentative phase diagram by taking the smallest \Rn\ values of sample A which still show a zero current state and the largest \Rn\ value which just shows a zero voltage state and estimate the \Es/\EL\ ratio according to the QPS theory. By assuming that this ratio is the same for other samples, the thick black lines in Fig.~\ref{phase_diagram} indicate the phase transition for nanowires with differences in length and in $R_\xi$ (normal-state resistance per coherence length). The shortest wire, sample C, shows no insulating phase, which is consistent with the phase diagram. For comparison, we added the data from Bollinger et al. \cite{Bollinger2008} together with the analysis from Mooij, et al. \cite{Mooij_2015}.  The measurement data of Bollinger et al. do not strictly distinguish between insulating and metallic, thus the data appear in Fig.~\ref{phase_diagram} in both the insulating and the metallic phase. For the studied wires, the transitions from  superconducting (red dots) to metallic (gray dots) and further up to insulating state (blue dots) nicely coincides with the solid lines drawn for constant $E_\mathrm{s} / E_\mathrm{L}$ ratios of $10^{-4}$ and $2\cdot10^{-2}$. Except for very short wires, the agreement is remarkable, especially if one takes the large parameter space, the very different material systems and employed techniques into account. 

\section{Conclusion and outlook}
In this paper, we have demonstrated that the normal-state resistance of narrow nanowires made from oxidized (granular) aluminum can be reduced in-situ at low temperatures by using current pulses. This tuning allowed us to observe a transition from insulating to superconducting behavior in a very controlled way, showing good agreement with theoretical models. We also report on the observation of a pronounced metallic behavior characterized by strongly non-linear conductance which is located between the insulating and the superconducting phase. The demonstrated intrinsic electromigration process (IEM) of the granular aluminum also provides a powerful new method of probing the superconductor to insulator transition. Using the QPS model we proposed a tentative phase diagram, predicting the transition parameters.

Apart from basic questions addressing superconducting nanowires, the results of this work may have impacts on more applied topics. The demonstrated strong non-linearity of the wires -- in conjunction with their low loss and their large kinetic inductance -- make them promising elements for superconducting quantum circuits \cite{Mooij_2005,Nazarov2011,Zorin2012,Grunhaupt2019,Schon2019}, qubits and metamaterials. Furthermore, particle detectors such as microwave kinetic inductance detectors or superconducting nanowire single-photon detectors may be optimised using the adjustability of the nanowire's resistance, as described in this work. 

\section{Methods}

\begin{small}
	
The nanowires are fabricated from a 20\,nm thick \Alox\ film deposited on a sapphire substrate. See Fig.~\ref{schematic} (b) for a scanning electron microscope image. The film has a sheet resistance of 2.7\,k$\Omega$ and $T_\mathrm{c} =  1.8$\,K. We use a single argon / chlorine based anisotropic etching step and a hydrogen silsesquioxane / polymethylmethacrylate bilayer etch mask to define the wire and the leads to the wire \cite{Rotzinger2016,Schon2019}. The wires have nominal width of 20\,nm.

The three measured nanowire samples are located on the same chip. Each wire has \Alox\ leads with a width of 0.5\,$\mathrm{\mu}$m and 2.5\,$\mathrm{\mu}$m length, that contribute to an inductance of $20$\,nH. With a stray capacitance of about $45$\,fF we estimate an environmental impedance of about 0.6\,k$\Omega$, not considering the inductance of the nanowire itself. 

The measurements were carried out in a dilution refrigerator at a temperature of 20\,mK. The electric bias leads are filtered at several stages from room temperature to the base temperature with copper powder, $\pi$ and RCR-low pass filters leading to an effective measurement bandwidth of about 5\,kHz. In the insulating regime, however, the $I-V$ characteristics are measured for a better signal to noise ratio with a minimal sampling time of about 0.02\,s. For the superconducting (low impedance) and insulating (high impedance) regimes, we use different amplifier readout schemes (see supplementary material for additional information). 

A computer-controlled measurement protocol was carried out for all samples in the following way: First, the $I-V$ characteristics were measured in either a voltage bias scheme (insulating regime) or a current biased scheme (metallic and superconducting regime). Then $I_\mathrm{p}$ was applied for about 20\,ms followed by at least a few seconds waiting time, to allow the samples to recover into thermal equilibrium. The nanowire resistance $R\mathrm{_n}$ was determined with an excitation current below $I_\mathrm{p}$. This then was followed by the next $I-V$ measurement cycle, and so on \cite{patent_tuning}.

The wire resistance is given by $R_\mathrm{n} = R_\mathrm{tot} - R_\mathrm{L} - R_\mathrm{th}$, where $R_\mathrm{tot}$ is the total value of the resistance measured, $R_\mathrm{L} = 26.5$\,k$\Omega$ (samples A and C), $R_\mathrm{L} = 39$\,k$\Omega$ (sample B) are the resistances of the on-chip leads connecting the nanowires. $R_L$ is in very good agreement with estimates using the \Alox\ sheet resistance and the geometry of the leads. 

We also recognised a 'thermal' resistance offset $R_\mathrm{th}$ at larger current bias values which is of the order of 15\,k$\Omega$ and can be explained by considering Joule heating \cite{baumans_thermal_2016}. 

In the insulating and metallic regime the resistances were determined from $I_\mathrm{p}$. In the superconducting regime however, $R_\mathrm{n}$ was extracted from the resistive slopes of the $I-V$ characteristics above $I_\mathrm{c}$ which conveniently allows to determine $R_\mathrm{L}$ and $R_\mathrm{th}$.

The effective coherence length $\xi$ was extracted from temperature dependent measurements of the upper critical magnetic field $H_\mathrm{c2}$ for several $\mathrm{\mu}$m wide \Alox\ wires with a sheet resistance ranging from 2.0\,k$\Omega$ to 5.1\,k$\Omega$. We found a constant $H_\mathrm{c2} (T=0) = 4.5\pm0.2$\,T$/\mu_0$, consistent with Ref.\@ \cite{HC2_Deutscher}. From this result, we get $\xi_\mathrm{n}^{0} = 8 \pm 0.4$\,nm, which is in good accordance with the $10$\,nm value quoted in Refs. \cite{Bachar_2015, Sonora_2019}. We take $\xi_n^{0}$ as a starting point for unaltered highly resistive nanowires. 

We assume furthermore, that the wire cross-section is sufficiently homogeneous, that the wire's \Es\ is not dominated by a narrow constriction in the nanowire. However, possible defects cannot be ruled out completely. In electron microscopy scans the wires appear smooth and uniform with an edge roughness of the order of 1-2\,nm.
	
\end{small}

\section{\label{sec:acknowledgment}{Acknowledgment}}
We thank L. Radtke and S. Diewald of the KIT Nanostructure Service Laboratory for the support concerning the sample fabrication. The work was funded by the Initiative and Networking Fund of the Helmholtz Association, the Helmholtz International Research School for Teratronics (JNV, YS) and the Landesgraduiertenfoerderung (LGF) of the federal state Baden Wuerttemberg (MW). Further support was provided by the Ministry of Education and Science of the Russian Federation in the framework of the Program to Increase Competitiveness of the NUST MISIS (contract No. K2-2020-017). JHC acknowledges the support of the Australian Research Council Centre of Excellence funding scheme (CE170100039) and the NCI National Facility through the National Computational Merit Allocation Scheme. 

\bibliography{bib}
\bibliographystyle{apsrev4-1}
\clearpage
\newpage

\section{Supplementary Material}
\maketitle

\subsection{Experimental details}

To ensure a low noise impact on the samples, also several noise-reducing devices were installed for all leads, going down to the sample at $\approx$ 20mK. First, the signals are filtered by $\pi$-filters at room temperature, then followed by RCR-low pass filters at the 4K stage of the refrigerator. The resulting measurement bandwidth is about 5\,kHz. Additionally, meter long copper powder filters are installed to suppress high frequency noise. 

\begin{figure}[h!]
	\begin{minipage}[c]{0.99\columnwidth}
		\includegraphics[width=\textwidth]{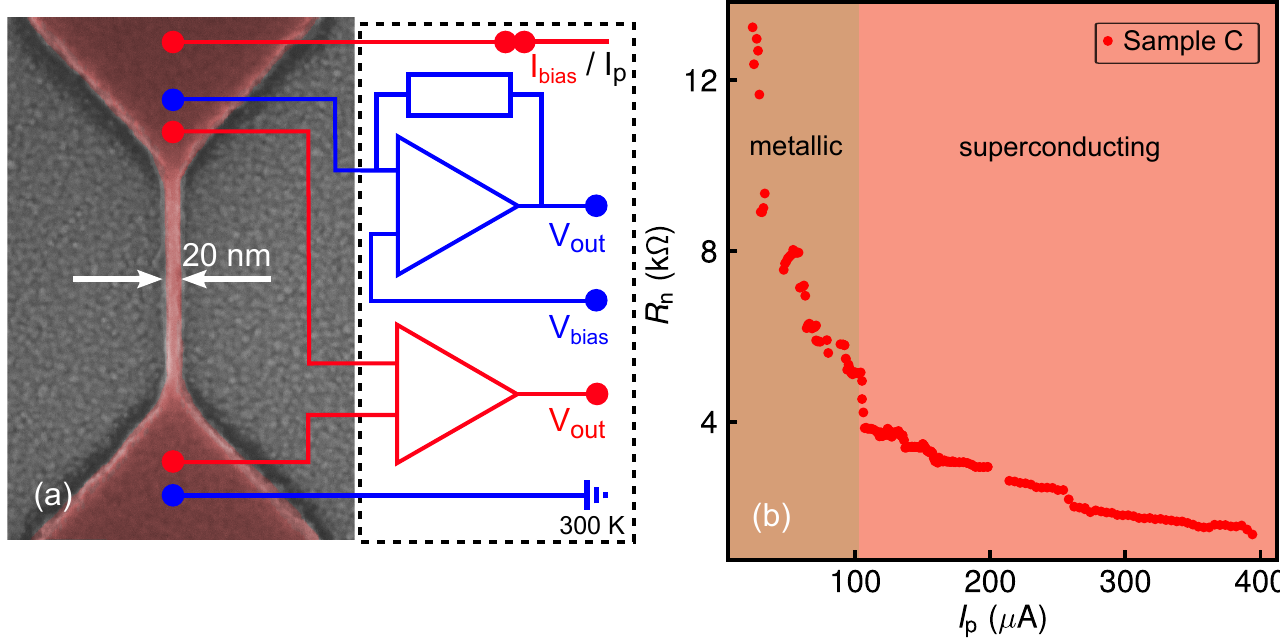}
	\end{minipage}
	\hfill
	\caption{\textbf{(a)} Measurement setup used to measure the $I-V$ characteristics and to reduce the normal-state resistance of the wire. Either the current bias scheme (red) or the voltage bias scheme (blue) was used. To automatise the procedure, we implemented several computer-controlled relays. $I_\mathrm{p}$ represents the applied pulse for the reduction of the wire resistance. \textbf{(b)} \Rn\ tuning for sample C as a function of the applied pulse current $I_\mathrm{p}$. The measurements were performed at 25\,mK. For small bias values, the electrical response was metallic (12.4\,-\,4.7\,k$\Omega$) and superconducting (4.7\,-\,1.5\,k$\Omega$).\label{measurement_setup}}
	
\end{figure}

A typical schematic diagram of the bias schemes we used is shown in Fig.~\ref{measurement_setup}: To measure and reduce the normal-state resistance of the wires and to record the $I-V$ characteristics in the metallic and superconducting regime, a current bias scheme was used (depicted in red). As current source ($I_\mathrm{bias}$, $I_\mathrm{p}$), we use a voltage-controlled, in-house made tunnel electronic which uses a Texas Instruments OPA2111 operational amplifier and current dividers as main elements. The input voltage signal is generated by a Keithley 2636A source meter and pre-filtered by a Stanford Instruments SR560 preamplifier. The output signal ($V_\mathrm{out}$) is amplified by a low-noise instrumentation amplifier INA 105KP and measured with the second channel of the source meter. In between the amplifier and the source meter, the signal is filtered by a Stanford Instruments SR560.

For samples A and B and being in the insulating regime, we used a voltage bias scheme (see Fig.~\ref{measurement_setup}, blue circuit). Here, one side of the sample is set to ground while on the other side, a FEMTO transimpedance amplifier (DDPCA-300) is applied. As for the current bias, the input voltage signal ($V_\mathrm{input}$) is generated by a Keithley 2636A and then filtered by a Stanford SR560. The output signal ($V_\mathrm{out}$) is measured in the same way as described before.  

\subsection{Coherence length}

\begin{figure}[h!]
	\begin{minipage}[c]{0.68\columnwidth}
		\includegraphics[width=\textwidth]{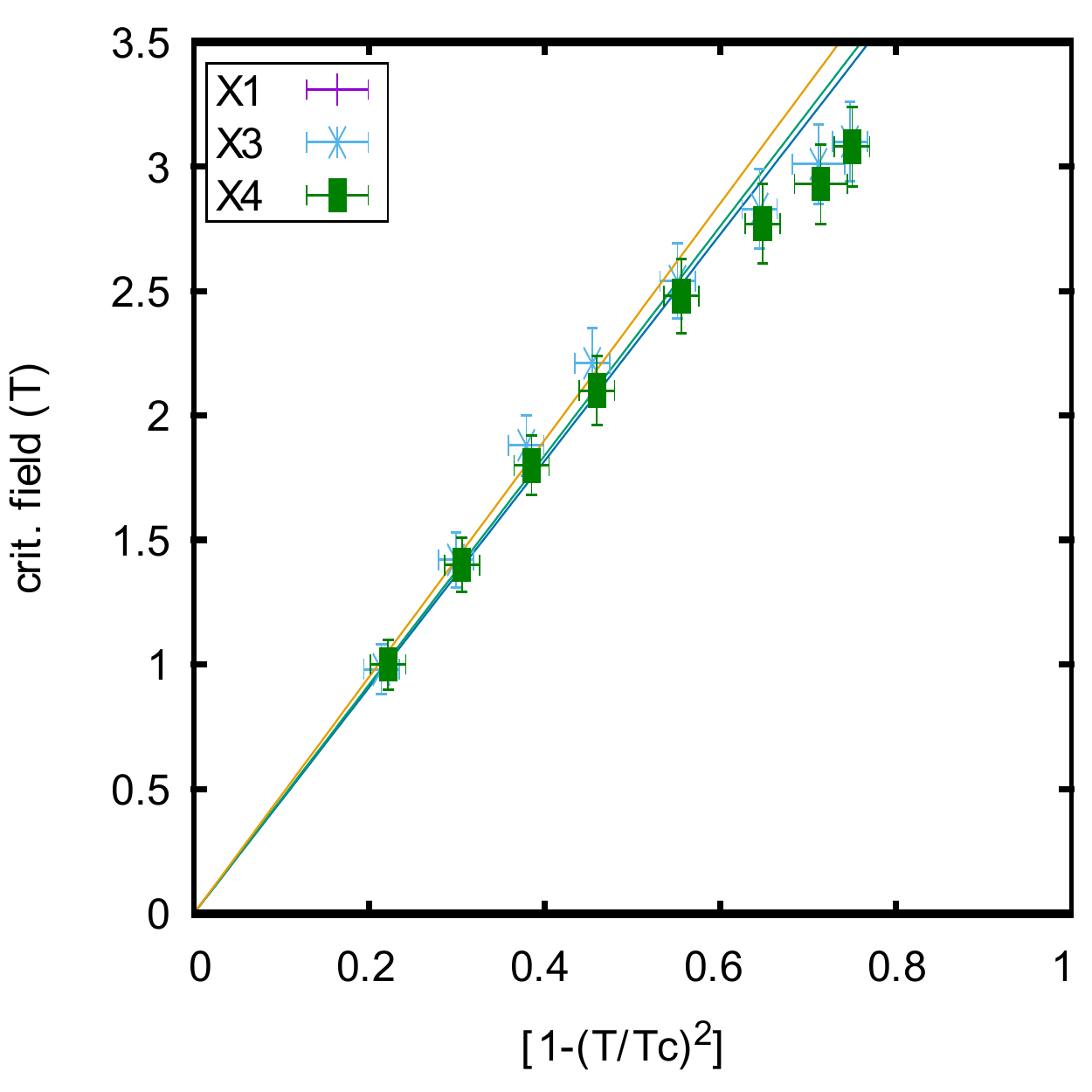}
	\end{minipage}	
	\hfill
	\caption{Dependence of the critical field $H_\mathrm{c2}$ on the reduced temperature $1-\left(T/T_\mathrm{c}\right)^2$. X1, X3 and X4 denote three different $\mathrm{\mu}$m wide \Alox\ wires. Sample X1 has a sheet resistance of $R_\mathrm{sq}^\mathrm{X1}=$\,2.0\,k$\Omega$, X3 of $R_\mathrm{sq}^\mathrm{X3}=$\,5.1\,k$\Omega$ and X4 of $R_\mathrm{sq}^\mathrm{X4}=$\,4.5\,k$\Omega$. The critical field $H_\mathrm{c2}$ is approximately the same for all samples at a certain temperature. \label{Hc2_measurements}
	}
	
\end{figure}

To determine the initial coherence length (before resistance tuning) for our samples, we have measured $H_\mathrm{c2} (T)$ for three $\mathrm{\mu}$m wide \Alox\ wires with different sheet resistances (see Fig.~\ref{Hc2_measurements}). For a type 2 superconductor, the relation between $H_\mathrm{c2} (T)$ and the Ginzburg Landau coherence length $\xi_\mathrm{GL}$ is given by: $H_\mathrm{c2} (T) = \Phi_0 /(2\pi \xi_\mathrm{GL}(T)^2)$ \cite{tinkham2004introduction}. Therefore $\xi_\mathrm{GL}(T=0)$ is directly related to $H_\mathrm{c2}(T=0)$. With the universal relation: $H_\mathrm{c2}(0) = 0.69 T_\mathrm{c} \left(dH_\mathrm{c2}/dT \right)_{T=T_\mathrm{c}}$ for a one-gap superconductor in the dirty limit (Werthammer at al., \cite{WerthamerHC2}), $H_\mathrm{c2}(0)$ can be extracted from the slope of $H_\mathrm{c2}(T)$ at $T=T_\mathrm{c}$. The critical temperatures were approximately 2\,K for all samples. From the linear fits to the $H_\mathrm{c2}(T)$ measurements (see Fig.~\ref{Hc2_measurements}), together with the measured $T_\mathrm{c}$ values, we receive a coherence length $\xi = 8 \pm 0.4$\,nm. 

\subsection{Critical currents}
For wires that are in the superconducting regime, we have compared the critical currents with the expectations of two different models (see Fig. \ref{FIG: 1um_supercond} (b)): Firstly, the model of Kulik and Omel'Yanchuk for a weak link in the dirty limit(KO1). Secondly, the model for a Josephson tunnel junction by Ambegaokar and Baratoff (AB). 

Both comparisons require a justification. In case of the AB model, we argue that, grossly simplified, the internal structure of the wires is very similar to a disordered chain of Josephson junctions where the weakest junctions are dominating $I_\mathrm{c}$. In contrast, the KO1 model considers the distributed character of a homogeneous weak link and is applicable up to an upper limit of the ratio wire / coherence length of about 3.49 which our wires exceed. We argue that, due to the internal granular structure, the comparison is still possible: The phase drop along the wire is not homogenous but instead it drops mainly over the oxide barriers between the nanoscale grains. It is therefore reasonable to introduce an effective length $L\mathrm{_{eff}}\simeq L_\mathrm{w}/5$, assuming an average grain size of 4\,nm and an oxide barrier of 1\,nm.

\subsection{Phase transitions}

\begin{figure}
	\begin{minipage}[c]{0.49\columnwidth}
		\includegraphics[width=\textwidth]{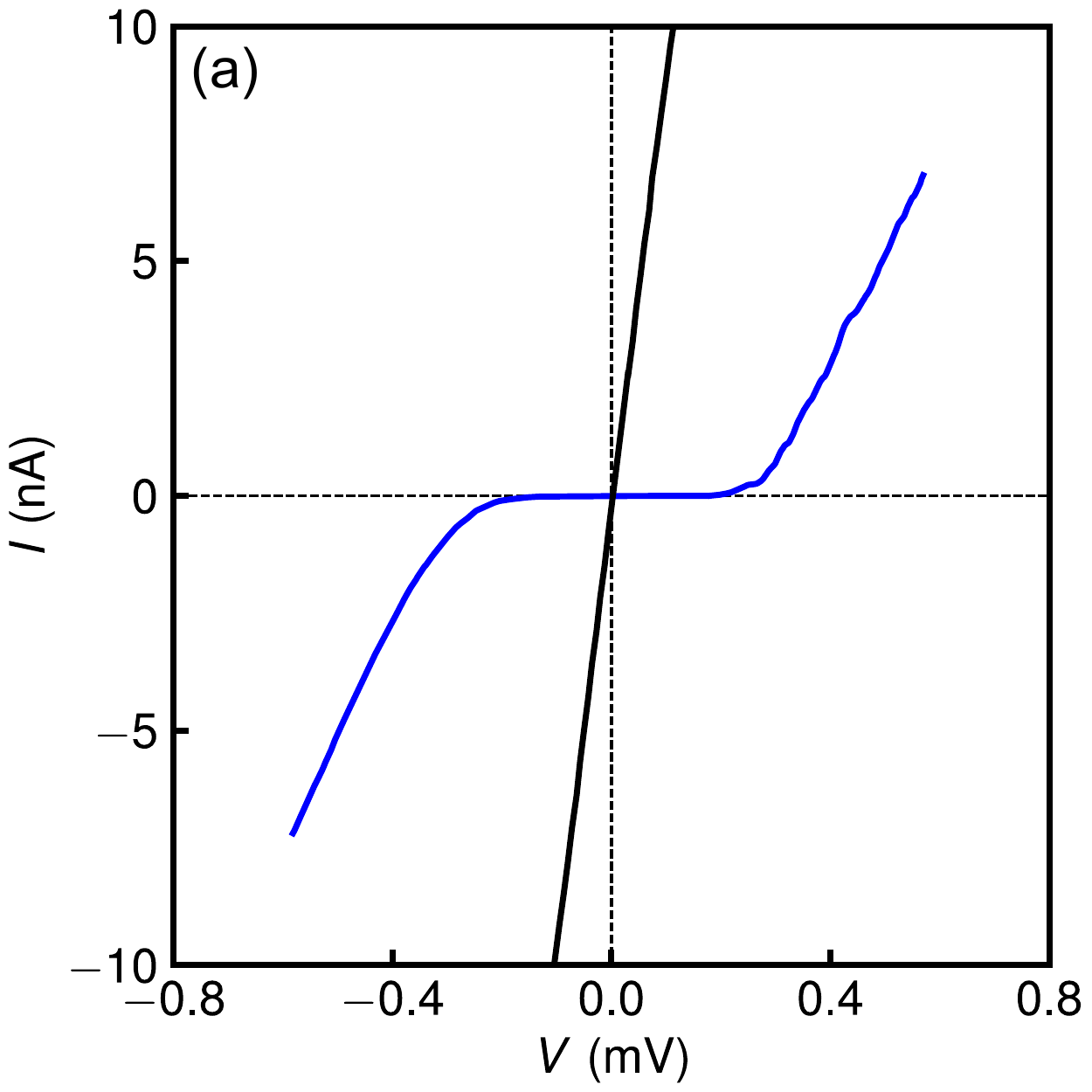}
	\end{minipage}
	\begin{minipage}[c]{0.49\columnwidth}
		\includegraphics[width=\textwidth]{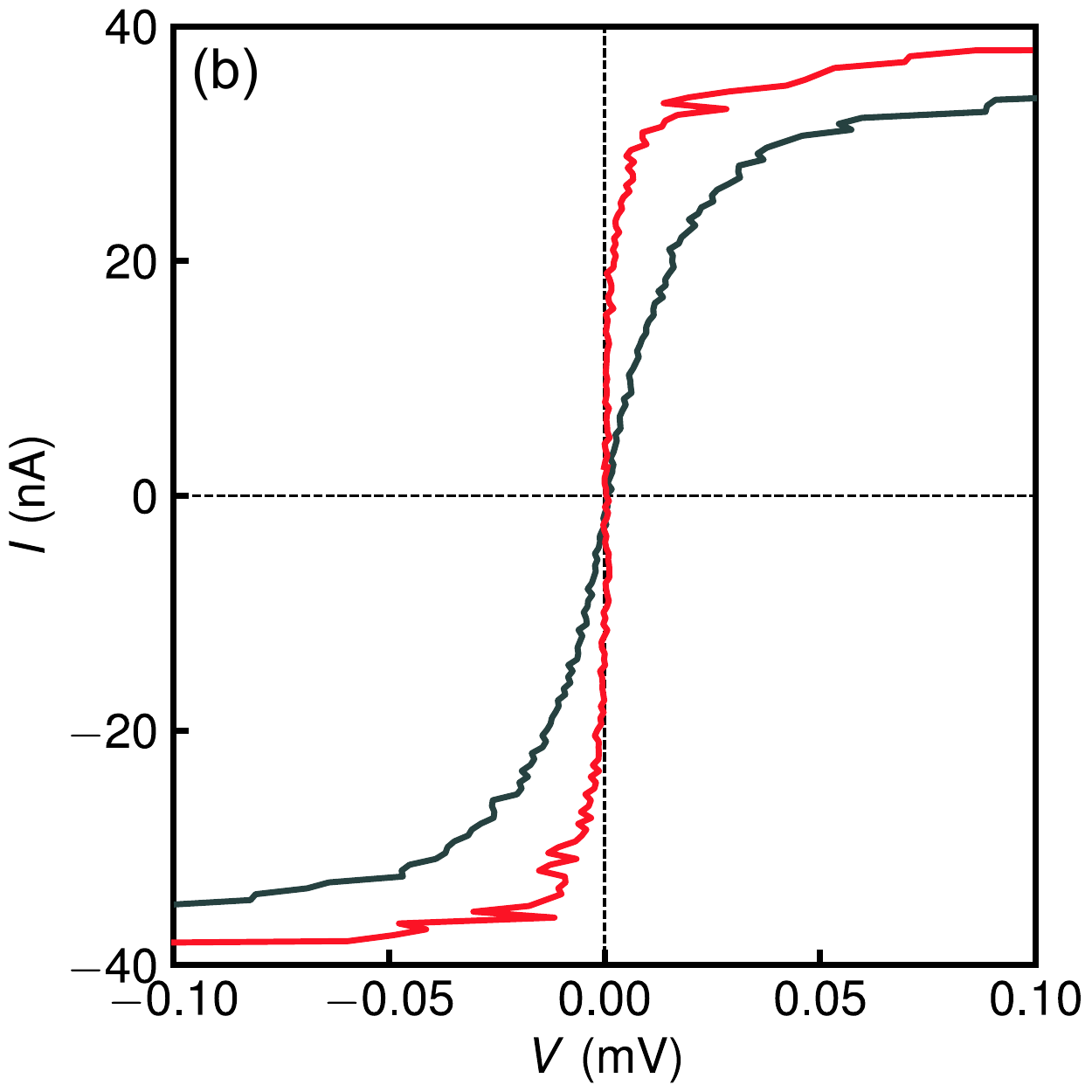} 
	\end{minipage}
	\hfill
	\caption{Change in the $I-V$ characteristics of sample A at the phase transition lines (see solid black lines in Fig. \ref{phase_diagram}). (a): For \Rn$>$ 50\,k$\Omega$, the response is insulating (blue line), below it is metallic (black line). (b) The second transition appears at \Rn$\approx$ 18\,k$\Omega$. Here the behavior again changes abruptly, but now from metallic (black line) to a superconducting behavior (red line).\label{transitions}}
\end{figure}

As described in the main part of this paper, at low temperatures and small bias values we distinguish between three different phases, namely insulating, metallic and superconducting. The phase diagram (Fig. \ref{phase_diagram}) reflects the transition through the different phases for all samples, driven by the intrinsic electromigration process (IEM). Even though it is possible to go in small resistance steps very smoothly through the different phases,  we observe a sharp border, separating the phases. Here, the $I-V$ characteristics change dramatically between two IEMs. Fig. \ref{transitions} shows this abrupt change in the transport behavior for sample A at both transitions, insulating-metallic (a), metallic-superconducting (b). 

\begin{figure}
	\begin{minipage}[c]{0.49\columnwidth}
		\includegraphics[width=\textwidth]{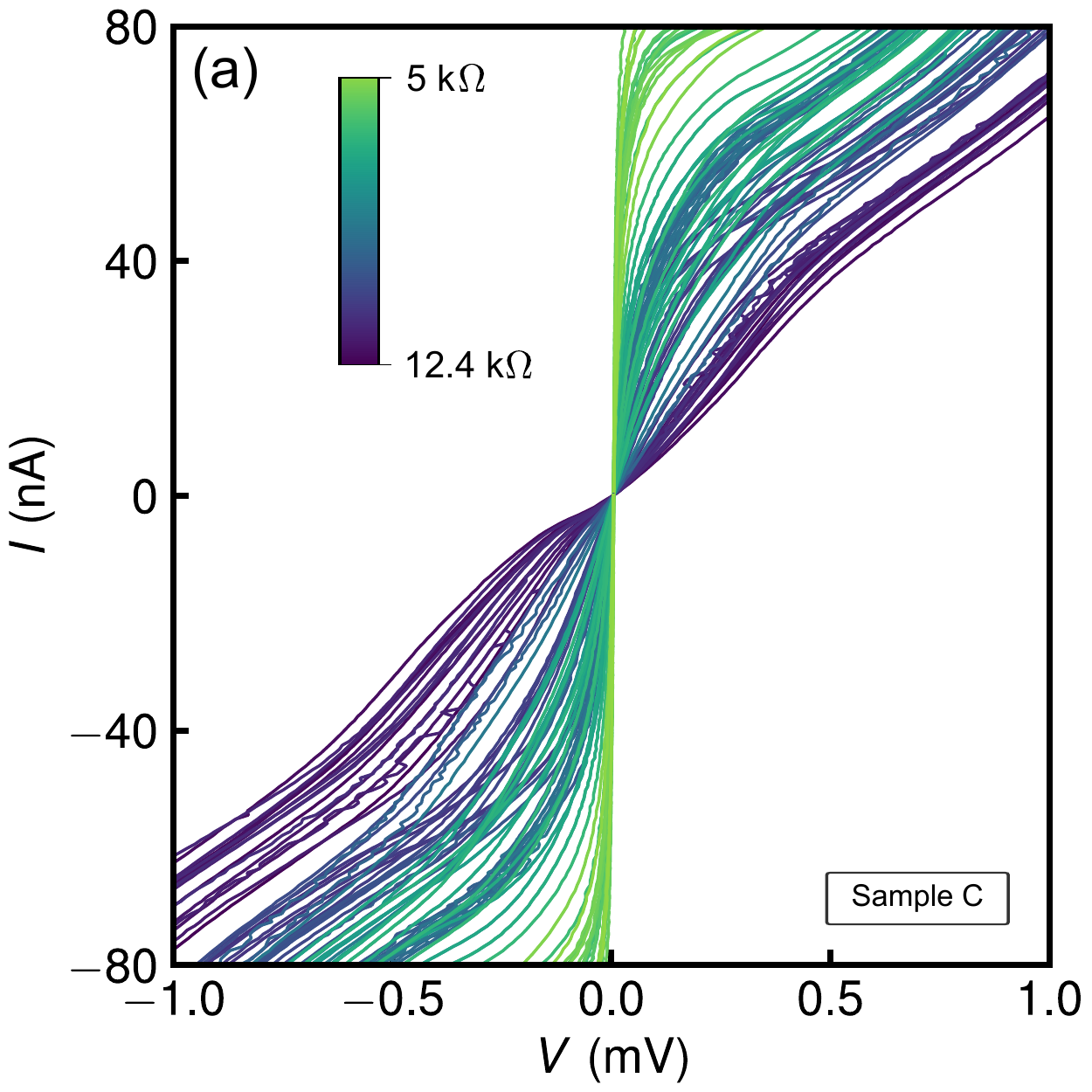}
	\end{minipage}
	\begin{minipage}[c]{0.49\columnwidth}
		\includegraphics[width=\textwidth]{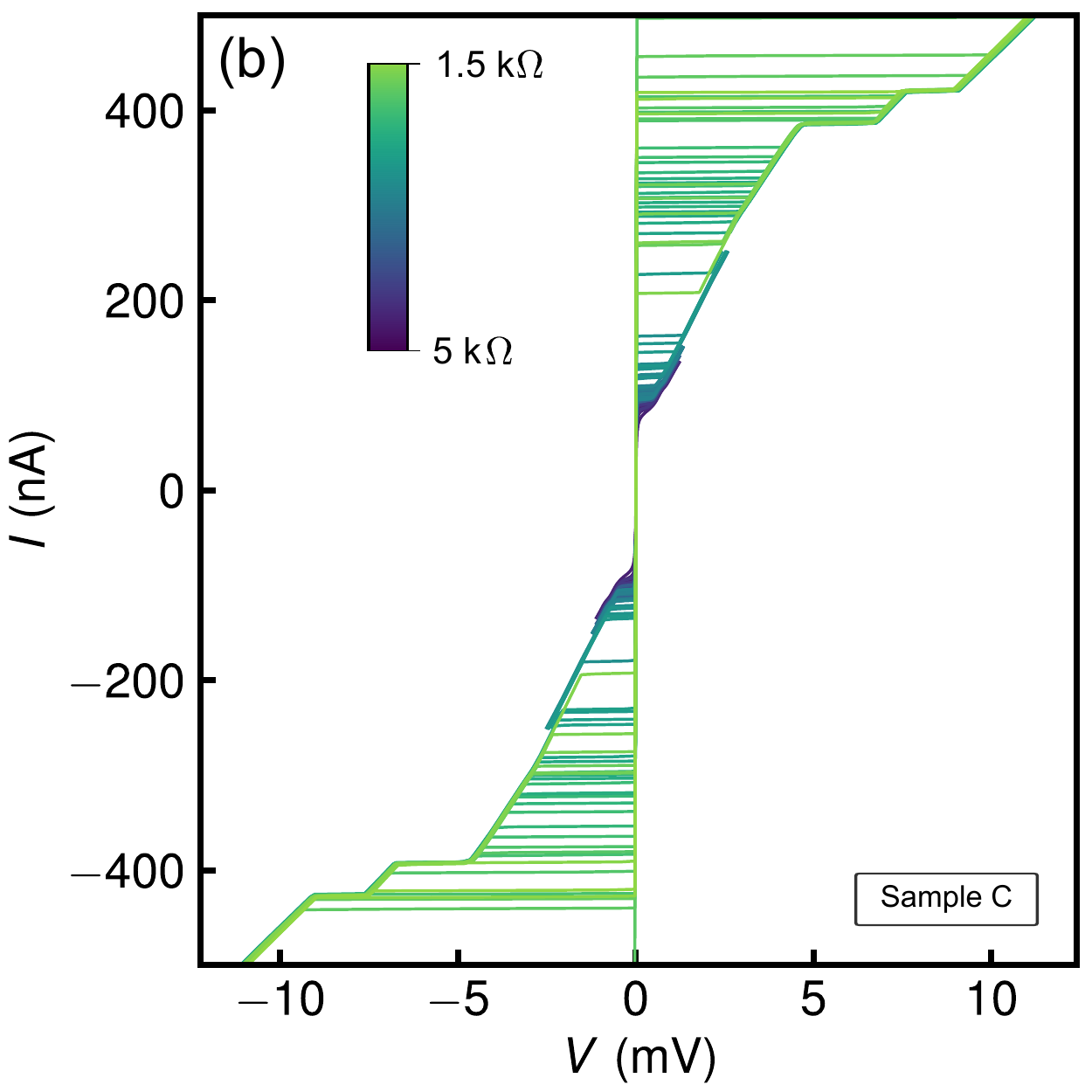} 
	\end{minipage}
	\hfill
	\caption{$I-V$ characteristics of sample C (250\,nm, for a detailed overview of the parameters see Table \ref{table}). \textbf{(a)} Metallic regime (12.4\,-\,4.7\,k$\Omega$): Sample C, in distinction from samples A and B, initially showed a metallic behavior. \textbf{(b)} Superconducting regime: For small bias currents, up to a critical value $I_\mathrm{c}$, no voltage drop is observed. The linear subbranches are caused by the on-chip leads. Here, the superconductivity breaks down sequentially in distinct steps.\label{IV250nm}}
\end{figure}

In contrast to samples A and B, sample C initially showed a metallic behavior (see Fig.~\ref{IV250nm} (a)). This is in good accordance with the reduced length, compared to the other samples, and the relatively small initial normal-state resistance \Rn\ of about 12\,k$\Omega$. The deviation of the measured \Rn\ from the expected normal-state resistance $R\mathrm{_{n}^{*}}>$\,33\,k$\Omega$, assuming a sheet resistance of 2.7\,k$\Omega$, can be addressed to a location-dependent variation of the film sheet resistance. However, for this sample, we were able to reduce $R_\mathrm{\xi}$ very smoothly through the metallic - superconducting transition illustrated in Fig.~\ref{phase_diagram}. The change in \Rn\ is shown in Fig. \ref{measurement_setup} (b). In the metallic regime, we have noted a substantial difference in the shape of the $I-V$ characteristics, depending on \Rn. For the highest resistances, the slope in the vicinity of zero bias increases with increasing bias current and therefore conductance. In contrast, for the smallest \Rn\ values the slope decreases monotonously with the increase of the bias current. At $R\mathrm{_n}\approx 5$\,k$\Omega$, the wire enters the superconducting regime with critical currents up to $\approx$\,440\,nA for the smallest \Rn\ (see Fig.~\ref{IV250nm} (b)).

\end{document}